\documentclass[pre,twocolumn,aps,groupedaddress]{revtex4}
\usepackage[T1]{fontenc}
\usepackage[latin9]{inputenc}
\usepackage{array}
\usepackage{booktabs}
\usepackage{mathrsfs}
\usepackage{mathbbol}
\usepackage{multirow}
\usepackage{amsmath}
\usepackage{amsthm}
\usepackage{amssymb}
\usepackage{stmaryrd}
\usepackage{graphicx}
\usepackage{latexsym}
\usepackage{textcomp}
\usepackage{mathtools}
\usepackage{xcolor}
\usepackage[citecolor=magenta,colorlinks=true]{hyperref}
\usepackage{lineno}
\usepackage{stackrel}
\usepackage[toc,page,titletoc]{appendix}
\usepackage{bm}
\definecolor{mycolor1}{rgb}{0.1, 0.6, 0.6}

\begin{document}

\title{Emergent power-law interactions in disordered crystals}

\author{Pappu Acharya}
\email{pappuacharya@tifrh.res.in}

\affiliation{Centre for Interdisciplinary Sciences, Tata Institute of Fundamental Research, Hyderabad 500107, India}

\author{Debankur Das}
\email{debankurd@tifrh.res.in}

\affiliation{Centre for Interdisciplinary Sciences, Tata Institute of Fundamental Research, Hyderabad 500107, India}

\author{Surajit Sengupta}

\affiliation{Centre for Interdisciplinary Sciences, Tata Institute of Fundamental Research, Hyderabad 500107, India}

\author{Kabir Ramola}
\email{kramola@tifrh.res.in}

\affiliation{Centre for Interdisciplinary Sciences, Tata Institute of Fundamental Research, Hyderabad 500107, India}

\begin{abstract}
We derive exact results for the fluctuations in energy produced by microscopic disorder in near-crystalline athermal systems. Our formalism captures the heterogeneity in the elastic energy of polydispersed soft disks in energy-minimized configurations. We use this to predict the distribution of interaction energy between two defects in a disordered background. 
We show this interaction energy displays an {\it average} power-law behaviour $\langle \delta E \rangle \sim \Delta^{-4}$ at large distances $\Delta$ between the defects.
These interactions upon disorder average also display the sixfold symmetry of the underlying reference crystal.
Additionally, we show that the fluctuations in the interaction energy encode the athermal correlations introduced by the disordered background. We verify our predictions with energy minimized configurations of polydispersed soft disks in two dimensions.
 
\end{abstract}


\maketitle

{\it Introduction:}
Emergent interactions arise in many microscopic theories of matter~\cite{gendelman2016emergent,steimel2016emergent,berezhiani2019emergent,qian2010hydrodynamic}.
A paradigmatic example is the Lennard-Jones interaction that emerges as an induced dipole effect of polarizable constituents \cite{lennard1931cohesion,london1937general, phillips2001crystals}.
Renormalization group techniques have allowed a precise calculation of many emergent effects from microscopic interactions. However, extending such techniques in the presence of disorder remains a theoretical challenge~\cite{ma1979random}.
An interesting example is particle interactions induced by an athermal embedding material, which arises in several natural contexts such as incompressible fluids and granular systems, and as yet have received less attention ~\cite{puljiz2016forces}. When cooled, many disordered systems display athermal behaviour \cite{grigera1999observation}, with temperature playing only a weak role in global properties, and represent a fundamentally different non-equilibrium material in comparison to liquids and gases.
Examples arise in various contexts in physics and biology and include systems displaying glassy behaviour \cite{berthier2011theoretical}, jammed packings of particles \cite{jaeger1996granular,van2009jamming,o2002random,o2003jamming,wyart2005rigidity,goodrich2012finite,ramola2017scaling,cates1998jamming, torquato2010jammed,bi2011jamming}, as well as densely packed tissues \cite{boromand2018jamming,broedersz2011criticality,dapeng1}. 
The local mechanical equilibrium constraints \cite{baule2018edwards,henkes2007entropy,bi2015statistical} lead to randomness at the microscopic scale, and such systems are not described by the usual elasticity theories in continuum \cite{landau1987theoretical,cui2019theory,biroli2016breakdown}, instead exhibiting emergent elasticity properties upon disorder average \cite{nampoothiri2020emergent}.

The elastic energy in amorphous athermal materials displays randomness, being sensitively dependent on the underlying disorder. The dynamics of such systems are governed primarily by the motion in phase space through energy minimized configurations. Characterizing the spatial distribution of the elastic energy as well as the fluctuations in the energy density produced by the underlying disorder is an important ingredient in any coarse-grained Hamiltonian-based elasticity theory.
Despite the importance and a number of previous studies \cite{sirote2021mean,golkov2019elastic,peyla1999elastic,peyla2003elastic} of the energy of interaction in such systems, an exact computation at the grain level has not been developed.
Moreover, continuum elasticity theories ignore the disorder present at the granular level \cite{peyla1999elastic}.
In this regard, exactly solvable models provide a fascinating arena to explore such questions. A prime candidate are near-crystalline jammed systems, where several exact predictions incorporating the effects of microscopic disorder can be made.
Near-crystalline jammed systems have recently been of interest, although they are relatively less studied \cite{Goodrich2014, tong2015crystals,acharya2020athermal,acharya2021disorder,tsekenis2021jamming,charbonneau2019glassy,das2021displacement}.
Indeed, the concept of jamming is not restricted to amorphous packings, with near-crystalline systems able to capture several non-trivial properties of jammed materials \cite{tong2015crystals,Goodrich2014}.




In this Letter, we present exact results for the fluctuation in energy produced by microscopic disorder in near-crystalline athermal systems. Our formalism predicts the {\it exact} energy of a system of soft disks with quenched disorder in the particle sizes. 
Using this, we derive the {\it average} interaction energy between defects placed at different locations in the system, which displays an emergent power-law behaviour at large distances. We also predict the exact distribution of the energy of interaction between two defects in the polydispersed background. 
Our results demonstrate that fluctuations in the interaction energy between defects encode the effects of the athermal disorder in the polydispersed backgrounds, a feature absent in thermal fluctuations.

{\it Disordered athermal crystals:}
\label{section_deformable_disks}
We study a system comprising of minimum energy configurations of soft disks interacting through the well-studied one-sided interaction \cite{durian1995foam,o2002random}
\begin{eqnarray}
\nonumber
U_{\sigma_{ij}}(\vec{r}_{ij}) &=& \frac{K}{\alpha}\left(1- \frac{| \vec{r}_{ij}|}{\sigma_{ij}}\right)^\alpha ~~\textmd{for}~~ r_{ij} < \sigma_{ij},\\
&=& 0 ~~~~~~~~~~~~~~~~~~~~~~\textmd{for}~~r_{ij} > \sigma_{ij}.
\label{energy_law}
\end{eqnarray}
Here $\vec{r}_{ij} = \vec{r}_{j} - \vec{r}_{i}$ is the vector distance between the $i^{th}$ and $j^{th}$ particles located at positions $\vec{r}_i$ and $\vec{r}_j$ respectively and $\sigma_{ij} = \sigma_i + \sigma_j$ is the sum of their individual radii. 
We set the effective stiffness of the interactions to $K = 1$ for simplicity. Although our results are valid for general $\alpha > 1$, we present results for the harmonic case ($\alpha = 2$). Notably, the {\it disorder} in this pairwise interaction is encoded in the quenched radii $\{ \sigma_{i} \}$. The interparticle forces are then given by $\vec{f}_{ij} = \frac{K}{\sigma_{ij}} \left(1- \frac{| \vec{r}_{ij}|}{\sigma_{ij}}\right)^{\alpha-1} \hat{r}_{ij}$,
where $\hat{r}_{ij}$ is the unit vector along 
the $\vec{r}_{ij}$ direction. We begin with a collection of equal sized soft disks with $\sigma_i = \sigma_0 = 1/2$. The minimum energy configuration is a crystalline state with the positions of the centers $\{\vec{r}_{i}^{(0)}\}$ forming a triangular lattice. The marginal crystal, with packing fraction $\phi_c = \pi/\sqrt{12}\approx 0.9069$, has no overlaps between particles, and zero interparticle forces. 
We work with overcompressed configurations, $\phi > \phi_c$ and the energy in the initial crystalline state $E^{(0)}_{ij}$ is equal at each bond $ij$. The total energy of the crystalline state is (see Supplemental Material \cite{supplemental})
\begin{equation}
E^{(0)} =\frac{1}{2}\sum_{i=1}^N \sum_{j=0}^5 E^{(0)}_{ij} = \frac{3 N}{2}\left(1-\sqrt{\frac{\phi_c}{\phi}}\right)^2,
\end{equation}
where $N$ is the number of particles in the system.
The superscript $(0)$ denotes the crystalline state with equal sized particles.
Next, we introduce disorder in the system by varying the radii of the particles as
\begin{eqnarray}
\sigma_i &=& (1 + \eta\xi_i)\sigma_0 = \sigma_0 +  \delta \sigma_i.
\end{eqnarray}
Here $\xi_i$ are independent identically distributed (i.i.d.) random variables. We choose a uniform underlying distribution of $\xi_i \in [-\frac{1}{2},\frac{1}{2}]$ \cite{tong2015crystals}, however our results are valid for any underlying distribution in the particle sizes.

\begin{figure}[t!]
\centering
\hspace*{-0.5cm}
\includegraphics[width=1.1\linewidth]{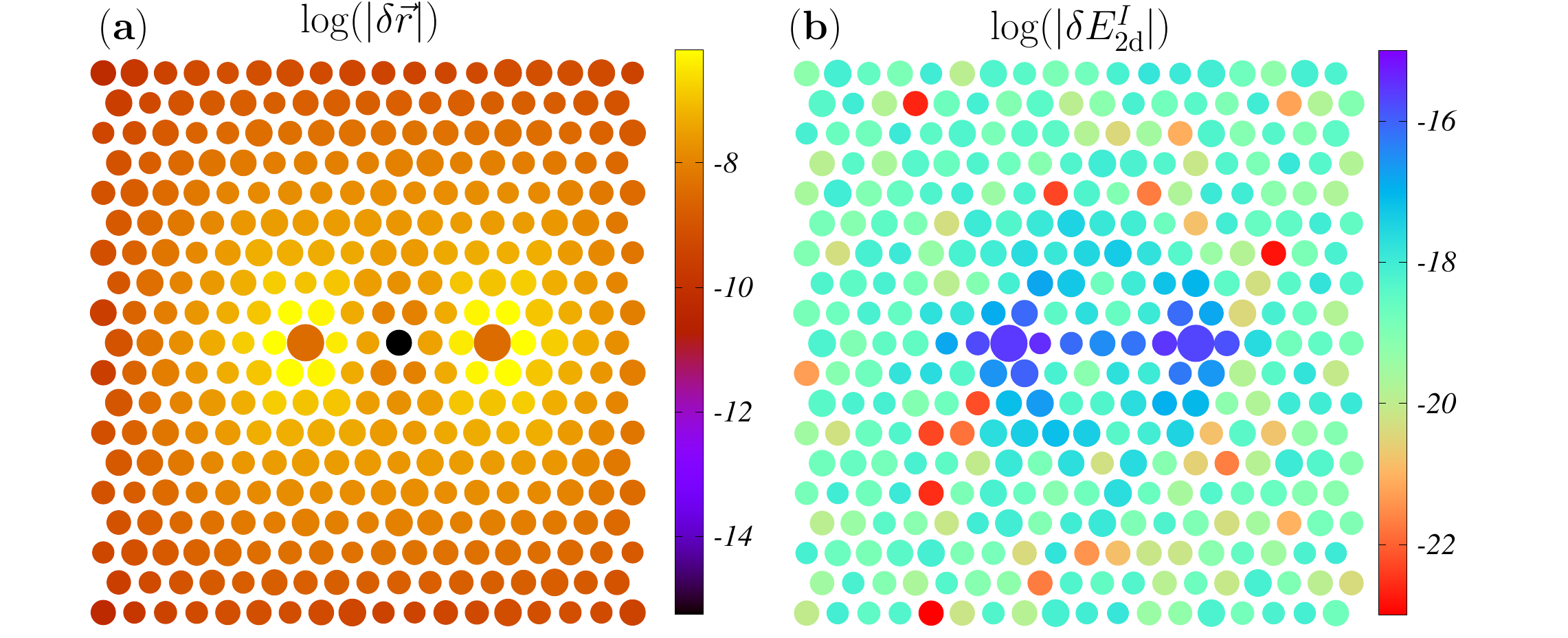}
\caption{
A section of an energy minimized configuration of a disordered jammed crystal with two defects. Disorder is introduced into the particle radii (with initial values $\sigma_i^{(0)} = 1/2$), with a polydispersity scale $\eta =5\times 10^{-6}$. The defect particles have an excess size $5\times10^{-3}$. {\bf (a)} The displacements ${\delta \vec{r}}$ from the crystalline positions are localized near the defects. {\bf (b)} The excess interaction energy $\delta E^I_{\text{2d}}$ between the defects is more heterogeneously distributed.
}
\label{schematic_fig}
\end{figure}

{\it Exact displacement fields:}
We begin by deriving exact displacement fields in energy minimized configurations with small disorder in particle sizes. For each realization of the quenched disorder, the system is allowed to relax into an energy minimized state, which is accomplished using the FIRE algorithm \cite{bitzek2006structural} in our numerical simulations. 
These configurations therefore satisfy the conditions of mechanical equilibrium i.e. $\sum_{j} f^{x}_{ij} = 0$, and $\sum_{j} f^{y}_{ij} = 0$, for each particle at site $i \equiv \vec{r}$. Here $f^{x(y)}_{ij}$ are the $x(y)$ components of the interparticle force between particles $i$ and $j$ in contact. Therefore, in order to characterize the behaviour of such a system, one needs to simultaneously solve {\it all} the force balance equations, which, along with the force-law yields a unique solution for particle displacements \cite{acharya2020athermal, acharya2021disorder}.
This can be accomplished with a systematic perturbation expansion about the crystalline ordered state, to linear order as well as higher orders \cite{acharya2020athermal,das2021long,acharya2021disorder}. When disorder is introduced into the system, the positions of the particles deviate from their crystalline values $\{ \vec{r}_{i}^{(0)} \} = \{x_{i}^{(0)}, y_{i}^{(0)}\}$ to a new mechanical equilibrium configuration $\{  \vec{r}_{i}^{(0)} + \delta \vec{r}_i \} = \{x_{i}^{(0)} + \delta x_i, y_{i}^{(0)}+\delta y_i\}$. The displacements can be expressed as a formal expansion
\begin{eqnarray}
\delta \vec{r}_i = \delta \vec{r}^{(1)}_{i} + \delta \vec{r}^{(2)}_{i} + \delta \vec{r}^{(3)}_{i} + \hdots,
\label{displacement_eq}
\end{eqnarray}
where $\{ \delta \vec{r}^{(n)}_{i} \} =  \{ \delta x^{(n)}_{i}, \delta y^{(n)}_{i} \}$ 
represent the $n^{th}$ order displacement fields of magnitude $\mathcal{O}(\eta^n)$. We focus on the terms upto second order $\delta x_{i} = \delta x^{(1)}_{i} + \delta x^{(2)}_{i}$ and $\delta y_{i} = \delta y^{(1)}_{i} + \delta y^{(2)}_{i}$, which contribute to the leading order terms in the energy of the system.
As the coefficients in the perturbation expansion are drawn from the underlying crystalline arrangement, the force balance equations can be solved hierarchically in Fourier space to obtain the displacement fields at every order \cite{acharya2021disorder}. The incremental radii $\{ \delta \sigma \}$ and the lower order displacement fields act as sources that generate the displacement fields at higher orders. At linear order, the displacements in Fourier space as a response to the disorder $ \{ \delta \sigma_{i} \}$ can be expressed as $\delta \tilde{ r}^{\mu,(1)}(\vec{k}) = \tilde{G}^{\mu}(\vec{k})\delta {\sigma}(\vec{k})$ \cite{acharya2021disorder}. 
Here $ \delta \tilde{\sigma}(\vec{k}) = \sum_{\vec{k}} e^{i\vec{k}\cdot \vec{r}} \delta \sigma(\vec{r})$, $\delta r^{\mu}$ refers to $\delta x$ and $\delta y$ for $\mu = x,y$ respectively, and $\tilde{G}^{\mu}(\vec{k})$ represent the response Green's functions in Fourier space.
The reciprocal lattice vectors for the triangular lattice arrangement are $\vec{k} \equiv (k_x,k_y) \equiv \left( \frac{2 \pi l}{2 L}, \frac{2 \pi m}{L} \right)$ \cite{horiguchi1972lattice}. 



{\it Energy of a disordered configuration:}
Importantly, the exact displacement fields can be used to compute the excess energy produced by an arbitrary configuration of excess radii over the crystalline background. 
The energy $(E_{ij}=U(\vec{r}_{ij}))$ of each bond $ij$ can be expressed as a perturbation expansion in the displacement fields as
\begin{eqnarray}
\nonumber
E_{ij} &=& E^{(0)}_{ij} + \sum_{\mu} e^{\mu}_{ij} \delta r^\mu_{ij} + \sum_{\mu}\sum_{\nu} e^{\mu \nu}_{ij} \delta r^\mu_{ij} \delta r^\nu_{ij}\\
&&~~~~~~~+ e^{\sigma}_{ij} \delta \sigma_{ij} + e^{\sigma \sigma}_{ij} \delta \sigma_{ij} \delta \sigma_{ij},
\label{en_expansion_main}
\end{eqnarray}
where the indices $\mu,\nu \equiv x,y$, and $\delta x(y)_{ij} = \delta x(y)_{j} -\delta x(y)_{i}$ represent the relative displacement fields up to all orders. The coefficients $e^{\mu}_{ij}$ and $e^{\mu \nu}_{ij}$ can be expressed purely in terms of the reference crystalline structure and are therefore translationally invariant, i.e. they do not depend on the site index $i$.
$E^{(0)}_{ij}$ represents the energy of each bond $ij$ in a pure crystal with no polydispersity in particle sizes. 
As the excess energy of the system incorporates second order terms, we group terms in Eq.~\eqref{en_expansion_main} that contribute up to second order. The excess energy of each bond above the crystalline value can then be expressed to leading order as
$
\delta E_{ij} =   E_{ij}- E^{(0)}_{ij} = T^{(1)}_{1,ij} + T^{(2)}_{1,ij} + T^{(1)}_{2,ij}.
$
We have explicitly
\begin{eqnarray}
\label{eq_T_expressions}
\nonumber
T^{(1)}_{1,ij}  &=&   e^x_{ij}\delta x^{(1)}_{ij} + e^y_{ij}\delta y^{(1)}_{ij} +e^{\sigma}_{ij}\delta \sigma_{ij},\\
T^{(2)}_{1,ij}  &=&   e^x_{ij}\delta x^{(2)}_{ij} + e^y_{ij}\delta y^{(2)}_{ij},\\
\nonumber
T^{(1)}_{2,ij}  &=&   e^{xx}_{ij} \delta x^{(1)}_{ij}\delta x^{(1)}_{ij} + e^{xy}_{ij} \delta x^{(1)}_{ij}\delta y^{(1)}_{ij} + e^{yy}_{ij} \delta y^{(1)}_{ij}\delta y^{(1)}_{ij}\\
\nonumber
&&+ e^{x\sigma}_{ij} \delta x^{(1)}_{ij}\delta \sigma_{ij}
+ e^{y\sigma}_{ij} \delta y^{(1)}_{ij}\delta \sigma_{ij} + e^{\sigma\sigma}_{ij} \delta \sigma_{ij} \delta \sigma_{ij}.
\end{eqnarray}
Above, the superscripts of $T$ represent the order the displacement field solution contributing to the energy, while subscripts represent the order of the energy expansion. For example, $T^{(1)}_{2,ij}$ represents the second-order term in the energy expansion using the linear-order solution to the displacement fields.
These terms are easily summed by taking Fourier transforms of the relevant fields.
Interestingly, the contribution from the term containing the second-order displacement fields is precisely zero i.e. $\sum_{ij} T^{(2)}_{1,ij}=0$ (see Supplemental Material \cite{supplemental}). The contributions from the other terms are respectively
\begin{small}
\begin{eqnarray}
\sum_{ij}T^{(1)}_{1,ij} = \gamma_1 \delta \tilde{\sigma}(0);~~
\sum_{ij} T^{(1)}_{2,ij} = \sum_{\vec{k}}\gamma_2(\vec{k}) \delta \tilde{\sigma} (\vec{k}) \delta \tilde{\sigma} (-\vec{k}),
\end{eqnarray}
\end{small}
where $\gamma_1=\sum_j 2 e^{\sigma}_{ij}$ and 
the exact expression for the interaction kernel in Fourier space $\gamma_2(\vec{k})$ is provided in the Supplemental Material \cite{supplemental}.
Interestingly this function displays the underlying crystalline symmetries as shown in Fig.~\ref{fig_gamma2_en} {\bf(a)}, which displacement correlations at linear order do not \cite{das2021long}.
Finally, grouping terms and dividing by $2$ to avoid double counting of bonds, the excess energy of an arbitrary configuration of defects is then
\begin{small}
\begin{eqnarray}
\label{energy_all_defect}
\delta E(\{ \delta \sigma_i \}) &=&\frac{1}{2}\big( \gamma_1 \delta \tilde{\sigma}(0)+ \sum_{\vec{k}}\gamma_2(\vec{k}) \delta \tilde{\sigma} (\vec{k}) \delta \tilde{\sigma} (-\vec{k})\big).
\label{eq_energy_expression}
\end{eqnarray}
\end{small}
We note that the above expression provides the {\it exact} energy of a disordered configuration, given the incremental sizes $\{ \delta \sigma_i \}$ of the particles.
\begin{figure}[t!]
\centering
\hspace*{-0.6cm}
\includegraphics[width=1.1\linewidth]{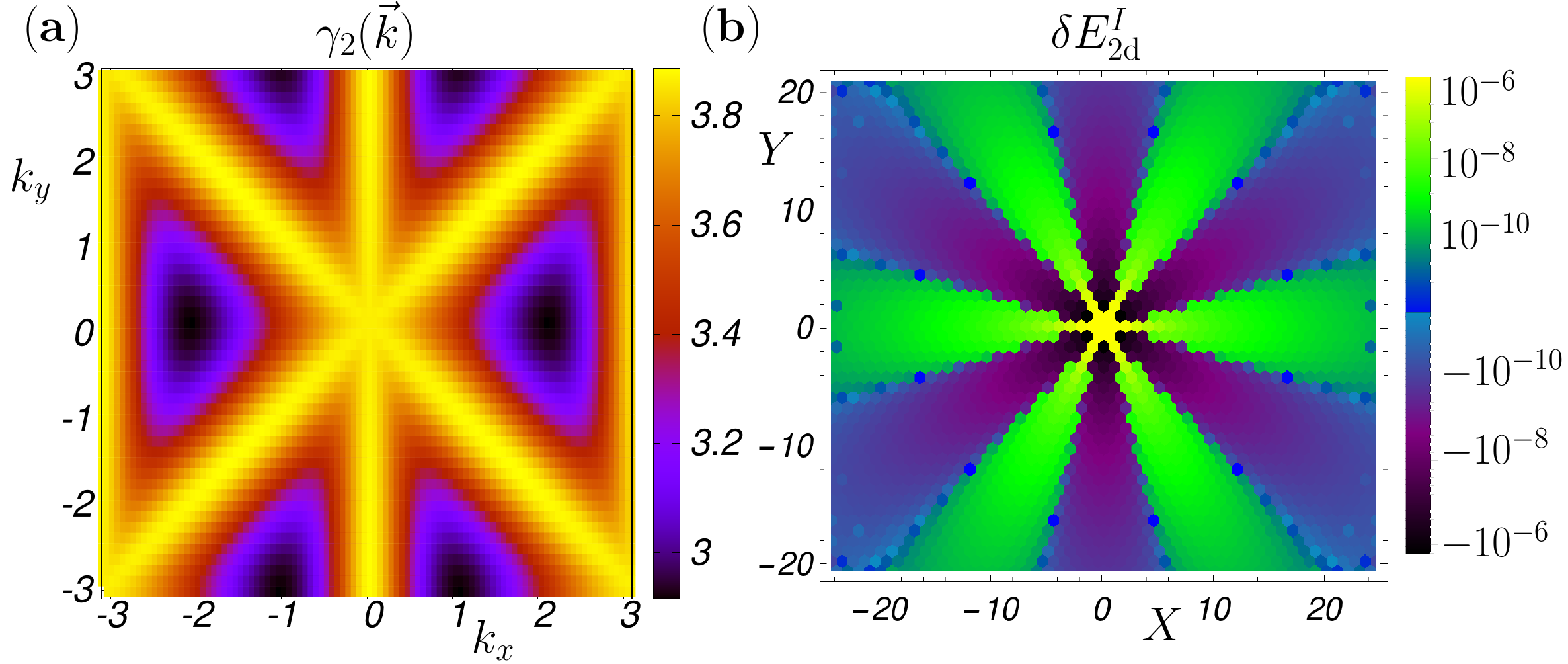}
\caption{{\bf (a)} Plot of the interaction kernel $\gamma_2(\vec{k})$ in Fourier space, displaying the underlying crystalline symmetries. {\bf (b)} The average energy of interaction between defects placed a distance $\vec{\Delta} \equiv (X,Y)$ apart. The average interaction energy displays positive and negative regions, reflecting the underlying crystalline background.
}
\label{fig_gamma2_en}
\end{figure}

{\it Disorder-averaged defect interactions:}
We next use our formalism to compute the energy of interaction between defects in a {\it disordered} background arising from polydispersity in particle sizes. We consider two particles with larger radii $\delta \sigma_1$ and $\delta \sigma_2$ in excess of the polydispersity scale, placed a relative distance $\vec{\Delta}$ apart. We place the defects at positions $\vec{0} = (0,0)$ and $\vec{\Delta} = (\Delta_x,\Delta_y)$, therefore $\delta \tilde{\sigma} (\vec{k})= \sum_{\vec{r}}\delta \sigma(\vec{r}) \exp{(\mathrm{i}\vec{k}.\vec{r})}=
\delta\sigma_{1}+ \delta\sigma_{2}\exp{(\mathrm{i}\vec{k}.\vec{\Delta})}$.
The energy of the system for the configuration with two defects can be expressed as $E = E^{(0)} + \delta E_{\text{1d}}(\vec{0}) + \delta E_{\text{1d}}(\vec{\Delta}) + \delta E^I_{\text{2d}}(\vec{0},\vec{\Delta})$, where each of the terms depend on the quenched disorder arising from the polydispersity $\{ \delta \sigma_i \}$. Here $\delta E_{\text{1d}}(\vec{0})$ and $\delta E_{\text{1d}}(\vec{\Delta})$ represent the excess energy associated with the defects $1$ and $2$ placed in the quenched background respectively, and $\delta E^I_{\text{2d}}(\vec{0},\vec{\Delta})$ represents the energy of interaction between the defects in the presence of the quenched disorder. 
We therefore have
\begin{eqnarray}
\hspace{-0.4cm}
\delta E^I_{\text{2d}}(\vec{0},\vec{\Delta}) &=& \delta E_{\text{2d}}(\vec{0},\vec{\Delta}) - \delta E_{\text{1d}}(\vec{0}) - \delta E_{\text{1d}}(\vec{\Delta}),
\label{eq_energy_terms}
\end{eqnarray}
The displacement fields as well as the interaction energy of a disordered configuration with two defects using the above expression are plotted in Fig.~\ref{schematic_fig}.
For large distances between defects ($|\vec{\Delta}| \to \infty$), $\delta E_{\text{2d}}(\vec{0},\vec{\Delta}) \to 0$, and the total energy of the system can be expressed as $E_{\infty} = E^{(0)} + \delta E_{\text{1d}}(\vec{0}) + \delta E_{\text{1d}}(\vec{\Delta})$.
Therefore, the energy of interaction between two defects can be obtained as $\delta E^I_{\text{2d}}(\vec{0},\vec{\Delta})  = E - E_{\infty}$. As $E_\infty$ is not accessible for finite system sizes, we consider the largest possible separation between defects. When the two defects are at a relative separation $\vec{\Delta}$, we choose $E_\infty \equiv E_{\vec{L}_{\text{max}}}$, where $\vec{L}_{\text{max}}$ represents the largest distance along $\vec{\Delta}$.
The energy of interaction then takes the form 
$\delta E^I_{\text{2d}}( \vec{\Delta}) = E_{\vec{\Delta}} - E_{\vec{L}_{\text{max}}}$.
This energy of interaction fluctuates for different realizations of the underlying disorder in particle sizes. Crucially, the fluctuations of $\delta E^I_{\text{2d}}$ in Eq.~\eqref{eq_energy_expression} are symmetric about the mean (see Supplemental Material \cite{supplemental} for details), leading to $\langle\delta E^I_{\text{2d}}\rangle=\delta E^I_{\text{2d}}(\eta = 0)$, where $\langle\rangle$ represents the disorder average over realizations, 
and $\delta E^I_{\text{2d}}(\eta = 0)$ represents the energy of the two defects in the crystalline background.
Using Eq.~(\ref{eq_energy_expression}) and Eq.~\eqref{eq_energy_terms} (see Supplemental Material \cite{supplemental}), the average energy of interaction for a finite system can be expressed as
\begin{small}
\begin{eqnarray}
\hspace{-0.0cm}
\nonumber
\langle \delta E^I_{\text{2d}}(\vec{0},\vec{\Delta}) \rangle  &=&  \delta\sigma_1 \delta\sigma_2 \sum_{\vec{k}}\gamma_2(\vec{k})\Big(\cos(\vec{k}.\vec{\Delta})-\cos \left(\vec{k}. \vec{L}_{\text{max}}\right)\Big),\\
\label{finite_ss_en}
\end{eqnarray}
\end{small}
The presence of the $\delta\sigma_1 \delta\sigma_2$ term in the above expression implies that the defects display both positive as well as negative interaction energies, depending on the positive or negative incremental sizes of the defects.
Additionally, as the interaction kernel $\gamma_2(\vec{k})$ possesses the crystalline symmetries, the {\it average} interaction energy also displays the sixfold symmetry of the underlying crystalline background, with positive and negative interactions along the lattice and off-lattice angles respectively.
We plot the disorder averaged interaction energy for arbitrary separations $\vec{\Delta} = (X,Y)$ in Fig.~\ref{fig_gamma2_en} {\bf (b)}.
Studies of induced dipole interactions in continuum elasticity have also revealed such an angular dependence of the interaction energy \cite{peyla1999elastic}. However, our results demonstrate that these interactions also emerge upon disorder average, with individual configurations displaying large heterogeneity, which we characterize below.

We next extract the asymptotic behaviour of the average interaction energy at large defect separations. We consider the infinite system size limit where the summation in Eq.~\eqref{finite_ss_en} can be converted to an integral. The average energy of interaction between two defects can then be expressed as
\begin{eqnarray}
\langle \delta E^I_{\text{2d}} \rangle = \frac{\delta\sigma_1 \delta\sigma_2 }{4\pi^2}\int_{-\pi}^\pi\int_{-\pi}^\pi\gamma_2(\vec{k})\cos(\vec{k}.\vec{\Delta})dk_x dk_y.
\label{en_allangla_nu}
\end{eqnarray}
The evaluation of this integral is rather involved and we provide the derivation in the Supplemental Material \cite{supplemental}. We focus on defects separated along the $x$-direction, and the final form of the interaction energy in the large distance limit ($\Delta \to \infty$) is given by
\begin{equation}
\langle \delta E^I_{\text{2d}} \rangle  \simeq  \mathcal{C} \frac{\delta\sigma_1 \delta\sigma_2}{\Delta ^4 }; ~~~ \mathcal{C} = \frac{24 \sqrt{3} (1-2 \epsilon )^2 (\epsilon -1)}{\pi (3-4 \epsilon )^2}.
\label{powerlaw}
\end{equation}
In Fig.~\ref{en}, we plot the numerically obtained interaction energies between two defects placed along the $x$-direction. 
We find that the average interaction predicted by our theory emerges as more realizations of the disorder are considered. 
We also demonstrate the convergence of the average interaction energy at large separations to the asymptotic power-law behaviour predicted in Eq.~(\ref{powerlaw}).
\begin{figure}[t!]
\hspace*{-1cm}
\includegraphics[scale=0.8]{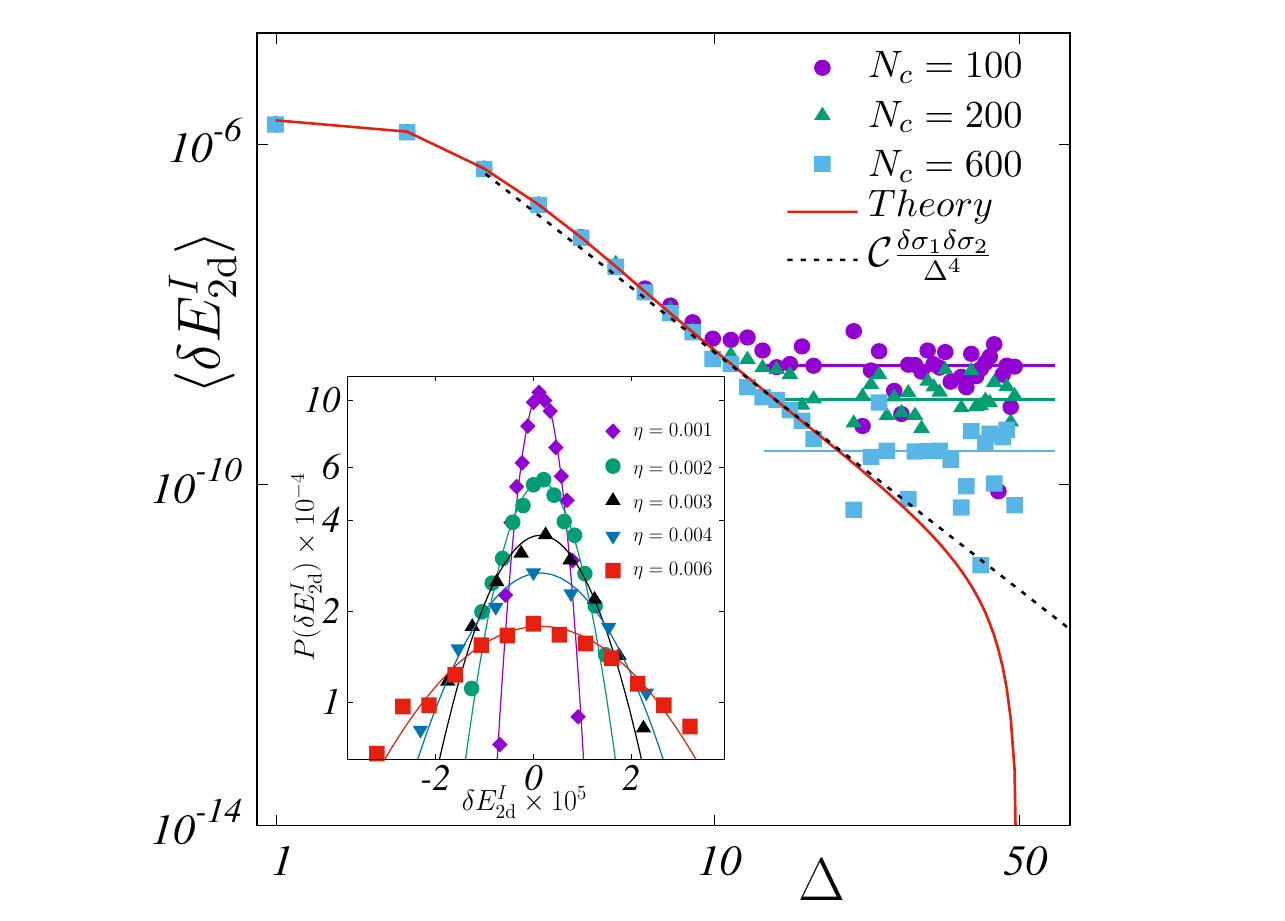}
\caption{The average interaction energy between two defects in a disordered background. 
The points represent data from simulations while the solid red line represents our theoretical prediction for the interaction energy in Eq.~\eqref{finite_ss_en}. The numerically obtained energy converges to the theoretical prediction with increasing number of configurations ($N_c$). The average energy displays an asymptotic $\sim\Delta^{-4}$ behaviour at large separations. {\bf (Inset)} Distribution of interaction energy between two defects placed at a distance of two lattice spacing apart for five different polydispersities. This energy is Gaussian distributed, and matches our predictions for the mean and variance in Eqs.~\eqref{finite_ss_en} and \eqref{eq_interaction_energy_fluctuation_main} exactly. Here we choose $\delta \sigma_1 = \delta \sigma_2 = 5 \times 10^{-3}$.
}
\label{en}
\end{figure}
{\it Fluctuations in interaction energy:} 
Finally, we turn our attention to the fluctuations in the interaction energy between defects, which can be used to characterize the emergence of the power-law interaction at large separations.
Using the expressions in Eqs.~(\ref{eq_energy_terms}) and (\ref{finite_ss_en}), the variance of the energy of interaction for an infinite system with two defects placed at a distance $\vec{\Delta}$ is given by (see Supplemental Material \cite{supplemental} for details)
\begin{eqnarray}
\label{eq_interaction_energy_fluctuation_main}
 \langle(\delta E_{\text{2d}}^I-\langle\delta E_{\text{2d}}^I\rangle)^2\rangle &=&\frac{V\eta^2}{48}\sum_{\vec{k}}
 \gamma_2(\vec{k})\gamma_2(-\vec{k})\\
 \nonumber
&& \left(\delta\sigma_1^2+\delta\sigma_2^2+\delta\sigma_1\delta\sigma_2 \cos(\vec{k}.\vec{\Delta})\right).
\end{eqnarray}
Interestingly, this variance in the interaction energy differs from the variance of the excess energies of the individual defects, as it depends on the distance $\vec{\Delta}$. 
This implies that the fluctuations in the interaction energy produced by two defects encodes the correlations produced by the microscopic disorder.
In the inset of Fig.~\ref{en}, we plot the distributions of the interaction energy along with the above theoretical predictions for five different $\eta$, displaying a near-exact match.  

{\it Discussion:}
In this Letter we have presented exact results for the fluctuations in energy of near-crystalline athermal systems. This was enabled by an exact characterization of the displacement fields for small disorder, which satisfies the microscopic force balance constraints. We used this framework to derive the {\it average} energy of interaction between defects in the presence of quenched disorder in particle sizes. Remarkably, the non-trivial power-law decay with distance $\sim \Delta^{-4}$ of continuum elasticity is recovered in this limit \cite{peyla1999elastic,peyla2003elastic}. Our results represent a microscopic derivation of this induced dipole effect in a disordered background. Moreover, our formulation allows us to predict the {\it fluctuations} in the energy of interaction, which matches our numerical results from energy minimized configurations of disordered athermal crystals exactly. The average energy of interaction displays an interesting angular dependence that encodes the symmetries of the underlying crystalline background. It would be interesting to study the effects of larger disorder in the system, in order to understand how rotational symmetry is recovered in the amorphous state, the nature of which continues to be the subject of intense scrutiny.

{\it Acknowledgments}: We thank Pinaki Chaudhuri, Silke Henkes, Bulbul Chakraborty, Subhro Bhattacharjee, Jishnu Nampoothiri, Vishnu V. Krishnan and Roshan Maharana for useful discussions. We acknowledge the contributions of Surajit Sengupta, now sadly deceased, to the initial part of this work. This project was funded by intramural funds at TIFR Hyderabad from the Department of Atomic Energy (DAE), Government of India.


\bibliography{Defect_Interactions_Bibliography.bib}

\clearpage

\begin{widetext}

 

\section*{\large Supplemental Material for\\ ``Emergent power-law interactions in disordered crystals''}

In this document we provide supplemental figures and details related to the results presented in the main text.

\maketitle

\subsection{Excess Energy of a Disordered Configuration}
In this Section we develop an exact theory to predict the excess energy produced by a general configuration of excess radii and use this to predict the interaction energy between defects. 
We evaluate the energy difference i.e., $\delta E = E -E^{(0)}$, where $E^{(0)}$ is the energy of the crystal (with no defects), and $E$ is the energy of the force balanced configuration with defects. The interaction potential between two particles is
\begin{eqnarray}
E_{ij}=U(\vec{r}_{ij}) = \frac{\epsilon}{2}\left(1- \frac{| \vec{r}_{ij}|}{\sigma_{ij}}\right)^2 =  \frac{\epsilon}{2}\left( 1-\frac{\sqrt{x^2_{ij}+y^2_{ij}}}{\sigma_{ij}}\right)^2, ~~\textmd{for}~~ r_{ij} < \sigma_{ij}.
\label{supp_energy_law}
\end{eqnarray}
The interparticle forces are defined as
\begin{equation}
\vec{f}_{ij} = \frac{\epsilon}{\sigma_{ij}} \left(1- \frac{| \vec{r}_{ij}|}{\sigma_{ij}}\right)^{\alpha-1} \hat{r}_{ij},
\label{supp_force_law_equation}
\end{equation}
where $\hat{r}_{ij}$ is the unit vector along the $\vec{r}_{ij}$ direction. 
For convenience we set $\epsilon = 1$.
Using Eqs.~(\ref{supp_energy_law}) and (\ref{supp_force_law_equation}), the total energy of the system in terms of the interparticle forces at each bond can be expressed as
\begin{eqnarray}
E&=&\frac{1}{4N}\sum_{i=1}^{N}\sum_{j=0}^{5}(\sigma_{ij}|f_{ij}|)^2.
\end{eqnarray} 
Here $i$ represents the particle index while $j$ represent its nearest neighbour as shown in Fig. \ref{fig_notation}.
\begin{figure}[ht!]
\centering
\includegraphics[width=0.3\linewidth]{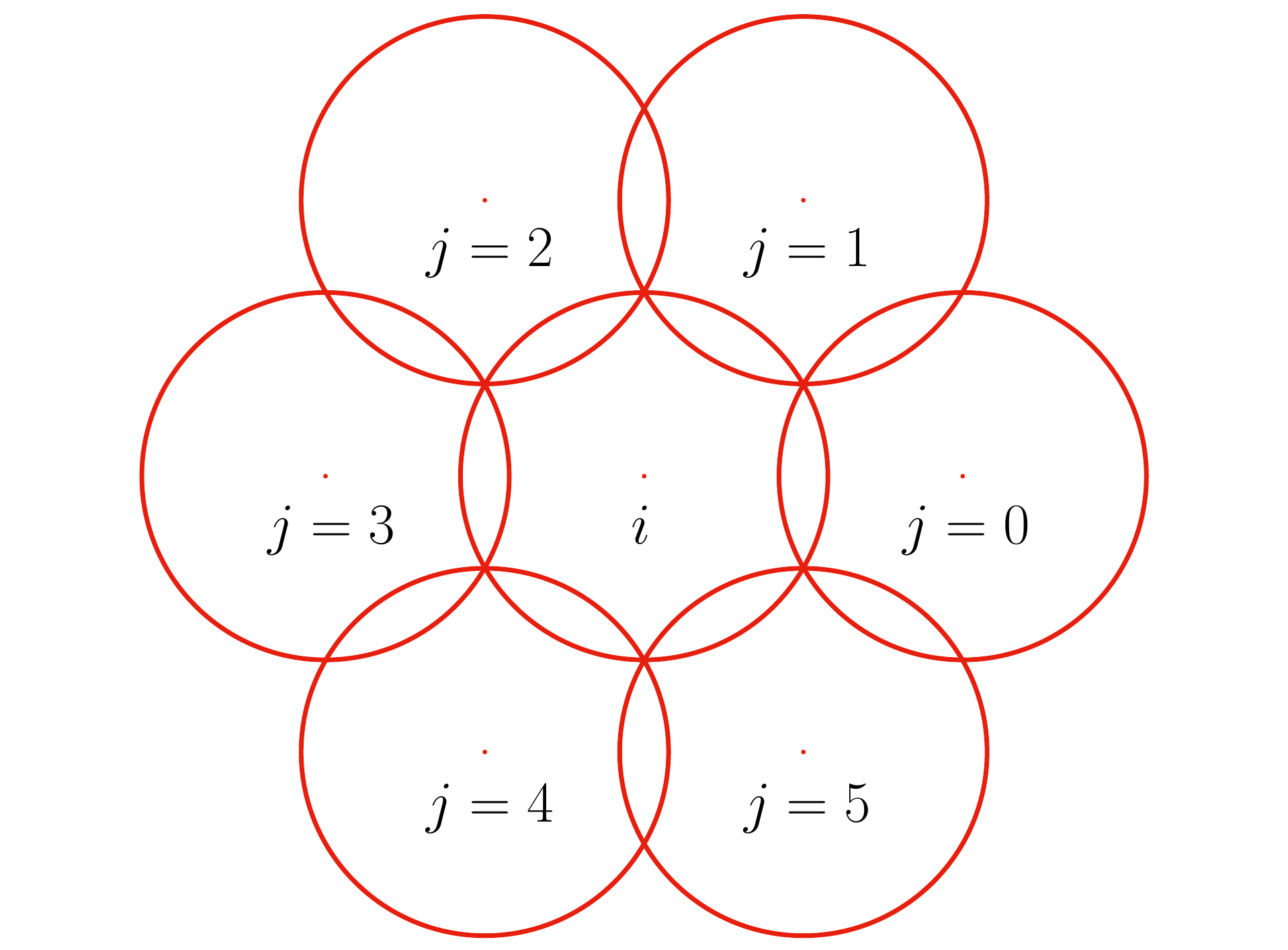}
\caption{The labeling convention for neighbours on the triangular lattice. Every particle $i$ has six nearest neighbours labeled $j=0$ to $5$ depending on their bond angles.}
\label{fig_notation}
\end{figure}

For the pure crystal with all radii equal, all the forces are equal in magnitude i.e. $|f_{ij}| = f_0$. The total energy per particle is given by
\begin{eqnarray}
E^{(0)}_{ij} &=& \frac{3}{2}(2\sigma_0)^2 |f_0|^2  =  \frac{3}{2}\left(1-\sqrt{\frac{\phi_c}{\phi}}\right)^2.
\end{eqnarray}


Perturbing about the crystalline ground state we can express the energy of each bond $ij$ as
\begin{eqnarray}
\begin{aligned}
E_{ij} &= \hspace{0.25 cm} E^{(0)}_{ij}\hspace{0.25 cm} + \hspace{0.25 cm}\Big( e^x_{ij}\delta x_{ij} + e^y_{ij}\delta y_{ij} +e^{\sigma}_{ij}\delta \sigma_{ij} + e^{xx}\delta x_{ij}\delta x_{ij}\\
& + e^{xy}_{ij}\delta x_{ij}\delta y_{ij}
+ e^{yy}_{ij}\delta y_{ij}\delta y_{ij}+ e^{x\sigma}_{ij} \delta x_{ij}\delta \sigma_{ij} + e^{y\sigma}_{ij} \delta y_{ij}\delta \sigma_{ij} + e^{\sigma\sigma}_{ij} \delta \sigma_{ij} \delta \sigma_{ij}\Big).
\end{aligned}
\label{en_expansion}
\end{eqnarray}
In the above expression the displacement fields $\delta x_{ij}$ represent the solutions at all orders i.e. $\delta x_{ij}=(\delta x_{ij}^{(1)}+ \delta x_{ij}^{(2)} + \hdots)$.
We consider the solutions of the displacement fields only up to second order, neglecting the higher order terms. 
We have from Eq.~(\ref{en_expansion})
\begin{eqnarray}
\delta E_{ij} &=&   E_{ij}- E^{(0)}_{ij}= T^{(1)}_{1,ij} + T^{(2)}_{1,ij} + T^{(1)}_{2,ij},
\end{eqnarray}
where the expressions for the different terms are provided in Eq.~(\ref{eq_T_expressions}) in the main text.
In order to calculate the excess energy of the crystal in the presence of an arbitrary configuration of defects, we need to sum over all bonds in real space. The term $T^{(1)}_{1,ij}$ when summed over all the bonds of the system yields
\begin{eqnarray}
\sum_{ij}T^{(1)}_{1,ij} &=& \sum_{ij} (e^x_{ij}\delta x^{(1)}_{ij} + e^y_{ij}\delta y^{(1)}_{ij} +e^{\sigma}_{ij}\delta \sigma_{ij}).
\label{t1}
\end{eqnarray}
The relative displacements in Fourier space can be expressed as
$\delta \tilde{r}^{\mu(1)}_{ij}(\vec{k}) = \left( 1- F_j(\vec{k})\right)\delta \tilde{r}^{\mu (1)}(\vec{k})$ and $\delta \tilde{\sigma}_{ij}(\vec{k}) = \left( 1+ F_j(\vec{k})\right)\delta \tilde{\sigma}(\vec{k})$. Here we define the basic translation coefficients in Fourier space 
\begin{eqnarray}
\mathcal{F}_j(\vec{k}) &=& \exp(- i \vec{k}.\vec{\mathbb{r}}_j),
\end{eqnarray}
where $\vec{\mathbb{r}}_j$ are the fundamental lattice translation vectors \cite{acharya2021disorder}, and can be written as $\vec{\mathbb{r}}_0 = (2,0)$, $\vec{\mathbb{r}}_1 = (1,1)$, $\vec{\mathbb{r}}_2 = (-1,1)$, $\vec{\mathbb{r}}_3 = (-2,0)$, $\vec{\mathbb{r}}_4 = (-1,-1)$, $\vec{\mathbb{r}}_5 = (1,-1)$.
Eq.~\eqref{t1} then leads to the following form 
\begin{eqnarray}
\nonumber
\sum_{ij}T^{(1)}_{1,ij} &=&\frac{1}{V}\sum_{ij}\sum_{\vec{k}}\Big( e^x_{ij} \left( 1- F_j(\vec{k})\right)\delta \tilde{x}^{(1)}(\vec{k})\exp{(-\mathrm{i} \vec{k}.\vec{r})} + e^y_{ij} \left( 1- F_j(\vec{k})\right)\delta \tilde{y}^{(1)}(\vec{k})\exp{(-\mathrm{i} \vec{k}.\vec{r})} \\
&&+ e^{\sigma}_{ij} \left( 1+ F_j(\vec{k})\right)\delta \tilde{\sigma}(\vec{k})\exp{(-\mathrm{i} \vec{k}.\vec{r})}\Big).
 \label{T1}
\end{eqnarray}
The derivations of the above equations and the form of the basic translation coefficients in Fourier space are provided in Ref.~\cite{acharya2021disorder}.
The linear order displacement fields in reciprocal space can be expressed as (see Ref.~\cite{acharya2021disorder} for details) 
\begin{eqnarray}
\delta \tilde{r}^{\mu}(\vec{k})&=&\tilde{G}^{\mu}(\vec{k})\delta\tilde{\sigma}(\vec{k}).
\end{eqnarray}
Next, using Eq.~\eqref{T1} and 
performing a summation over all sites $i$, we arrive at
\begin{eqnarray}
\sum_{ij}T^{(1)}_{1,ij} = \frac{1}{V}\sum_{j}\sum_{\vec{k}}\delta(\vec{k})\Big(e^x_{ij}\left( 1- F_j(\vec{k})\right)\tilde{G}^x(\vec{k})\delta \tilde{\sigma}(\vec{k}) + e^y_{ij}\left( 1- F_j(\vec{k})\right)\tilde{G}^y(\vec{k})\delta \tilde{\sigma}(\vec{k})
+ e^{\sigma}_{ij}\left( 1+ F_j(\vec{k})\right)\delta \tilde{\sigma}(\vec{k})\Big). 
\end{eqnarray}
Here $\delta(\vec{k})$ is the Kronecker delta function. Performing the summation over all $\vec{k}$ we arrive at 
\begin{eqnarray}
\sum_{ij}T^{(1)}_{1,ij} &=& \sum_j\delta \tilde{\sigma}(0)\Big( e^x_{ij} \left( 1- F_j(0)\right)\tilde{G}^x(0) + e^y_{ij} \left( 1- F_j(0)\right)\tilde{G}^y(0)  + e^{\sigma}_{ij} \left( 1+ F_j(0)\right)\Big).
\label{t11} 
\end{eqnarray}
Here $F_j(0)$ is 1. 
The displacement fields along the $x$ and $y$ directions produced by a single defect satisfy the following symmetry properties
\begin{eqnarray}
\nonumber
\delta x^{(1)}(x,y) &=& -\delta x^{(1)}(-x,y),\,\,\,\delta y^{(1)}(x,y) = -\delta y^{(1)}(x,-y).
\end{eqnarray}   
This indicates that for a system with a single defect, the summation of the displacement fields $\sum_i\delta x^{(1)}_i = 0$ and $\sum_i\delta y^{(1)}_i = 0$. The superposition principle at linear order dictates that the above relations are true for any arbitrary defect configuration.
Therefore, for any configuration of defects $\{\delta \sigma \}$, we obtain $\tilde{G}^\mu(\vec{k}) \delta \tilde{\sigma}(\vec{k}) = 0$ and hence we have
\begin{eqnarray}
\tilde{G}^\mu(0) &=& 0.
\end{eqnarray}
The term in Eq.~(\ref{t11}) therefore reduces to
\begin{eqnarray}
\sum_{ij}T^{(1)}_{1,ij} &=& \gamma_1 \delta \tilde{\sigma}(0), 
\end{eqnarray}
where 
$\gamma_1=\sum_j 2 e^{\sigma}_{ij}$.
Next, the first-order term of the excess energy derived from the second-order displacement fields can be expressed as
\begin{eqnarray}
\sum_{ij} T^{(2)}_{1,ij} &=& \sum_{ij} (e^x_{ij}\delta x^{(2)}_{ij} + e^y_{ij}\delta y^{(2)}_{ij})
= \sum_j\Big( e^x_{ij} \left( 1- F_j(0)\right)\tilde{G}^x(0) \tilde{S}^x(0) + e^y_{ij} \left( 1- F_j(0)\right)\tilde{G}^y(0)\tilde{S}^y(0)\Big). 
\end{eqnarray}
Here $\tilde{S}^x(\vec{k})$ and $\tilde{S}^y(\vec{k})$ are the second order source terms in the perturbation expansion (see Ref.~\cite{acharya2021disorder}) along the $x$ and $y$ directions respectively. 
As $\tilde{G}^x(0)$ and $\tilde{G}^y(0)$ are precisely zero, the term
$\sum_{ij} T^{(2)}_{1,ij}$ vanishes. Finally, the term $\sum_{ij} T^{(1)}_{2,ij}$ can be expressed as
\begin{eqnarray}
\sum_{ij} T^{(1)}_{2,ij} &=&\sum_{ij} \big(e^{xx}_{ij}\delta x^{(1)}_{ij}\delta x^{(1)}_{ij} + e^{xy}_{ij}\delta x^{(1)}_{ij}\delta y^{(1)}_{ij}
+ e^{yy}_{ij}\delta y^{(1)}_{ij}\delta y^{(1)}_{ij}+ e^{x\sigma}_{ij}\delta x^{(1)}_{ij}\delta \sigma_{ij} + e^{y\sigma}_{ij}\delta y^{(1)}_{ij}\delta \sigma_{ij}+ e^{\sigma\sigma}_{ij}\delta \sigma_{ij}\delta \sigma_{ij}\big).
\end{eqnarray}
The first term in the above expression can be expressed as
\begin{small}
\begin{eqnarray}
\sum_{ij} e^{xx}\delta x^{(1)}_{ij}\delta x^{(1)}_{ij} &=& \sum_{ij} e^{xx}\frac{1}{V}\sum_{\vec{k}}\left( 1- F_j(\vec{k})\right)\delta \tilde{x}^{(1)}(\vec{k})\exp{(-\mathrm{i} \vec{k}.\vec{r}_i)} 
\times
\frac{1}{V}\sum_{\vec{k}^\prime}\left( 1- F_j(\vec{k}^\prime)\right)\delta \tilde{x}^{(1)}(\vec{k}^\prime)\exp{(-\mathrm{i} \vec{k}^\prime.\vec{r}_i)}.
\end{eqnarray}
\end{small}
Summing over all bonds $ij$ yields
\begin{eqnarray}
\sum_{ij} e^{xx}_{ij} \delta x^{(1)}_{ij}\delta x^{(1)}_{ij} 
= \frac{1}{V} \sum_{j} e^{xx}_{ij} \sum_{\vec{k}}\left( 1- F_j(\vec{k})\right) \left( 1- F_j(-\vec{k})\right)\tilde{G}^x(\vec{k})\tilde{G}^x(-\vec{k})\delta \tilde{\sigma} (\vec{k}) \delta \tilde{\sigma} (-\vec{k}).
\end{eqnarray}
The rest of the terms contributing to the final expression of $\sum_{ij} T^{(1)}_{2,ij}$ can be treated in a similar manner, and we arrive at
\begin{eqnarray}
\nonumber
\sum_{ij} T^{(1)}_{2,ij} &=& \sum_{\vec{k}}\gamma_2(\vec{k}) \delta \tilde{\sigma} (\vec{k}) \delta \tilde{\sigma} (-\vec{k}),
\end{eqnarray}
where
\begin{eqnarray}
\nonumber
\gamma_2(\vec{k})&=&\frac{1}{V}\sum_{j}\Big( e^{xx} \left( 1- F_j(\vec{k})\right) \left( 1- F_j(-\vec{k})\right)\tilde{G}^x(\vec{k})\tilde{G}^x(-\vec{k}) +  e^{xy} \left( 1- F_j(\vec{k})\right) \left( 1- F_j(-\vec{k})\right)\tilde{G}^x(\vec{k})\tilde{G}^y(-\vec{k})\\
\nonumber
&&+  e^{yy} \left( 1- F_j(\vec{k})\right) \left( 1- F_j(-\vec{k})\right)\tilde{G}^y(\vec{k})\tilde{G}^y(-\vec{k}) +  e^{x\sigma} \left( 1+ F_j(\vec{k})\right) \left( 1- F_j(-\vec{k})\right)\tilde{G}^x(\vec{k})\\
&&+  e^{y\sigma} \left( 1- F_j(\vec{k})\right) \left( 1+ F_j(-\vec{k})\right)\tilde{G}^y(\vec{k}) + e^{\sigma \sigma} \left( 1+ F_j(\vec{k})\right) \left( 1+ F_j(-\vec{k})\right)\Big).
\label{gamma2_exact}
\end{eqnarray}


Grouping all the terms and dividing by $2$ to avoid double counting of bonds, the excess energy for an arbitrary configuration of defects can be expressed as
\begin{eqnarray}
\delta E = \frac{1}{2}\sum_{ij}\delta E_{ij} &=&\frac{1}{2}\big( \gamma_1 \delta \tilde{\sigma}(0)+ \sum_{\vec{k}}\gamma_2(\vec{k}) \delta \tilde{\sigma} (\vec{k}) \delta \tilde{\sigma} (-\vec{k})\big),
\end{eqnarray}
 which is Eq.~(\ref{eq_energy_expression}) in the main text.
\begin{figure}[t!]
\includegraphics[scale=0.80]{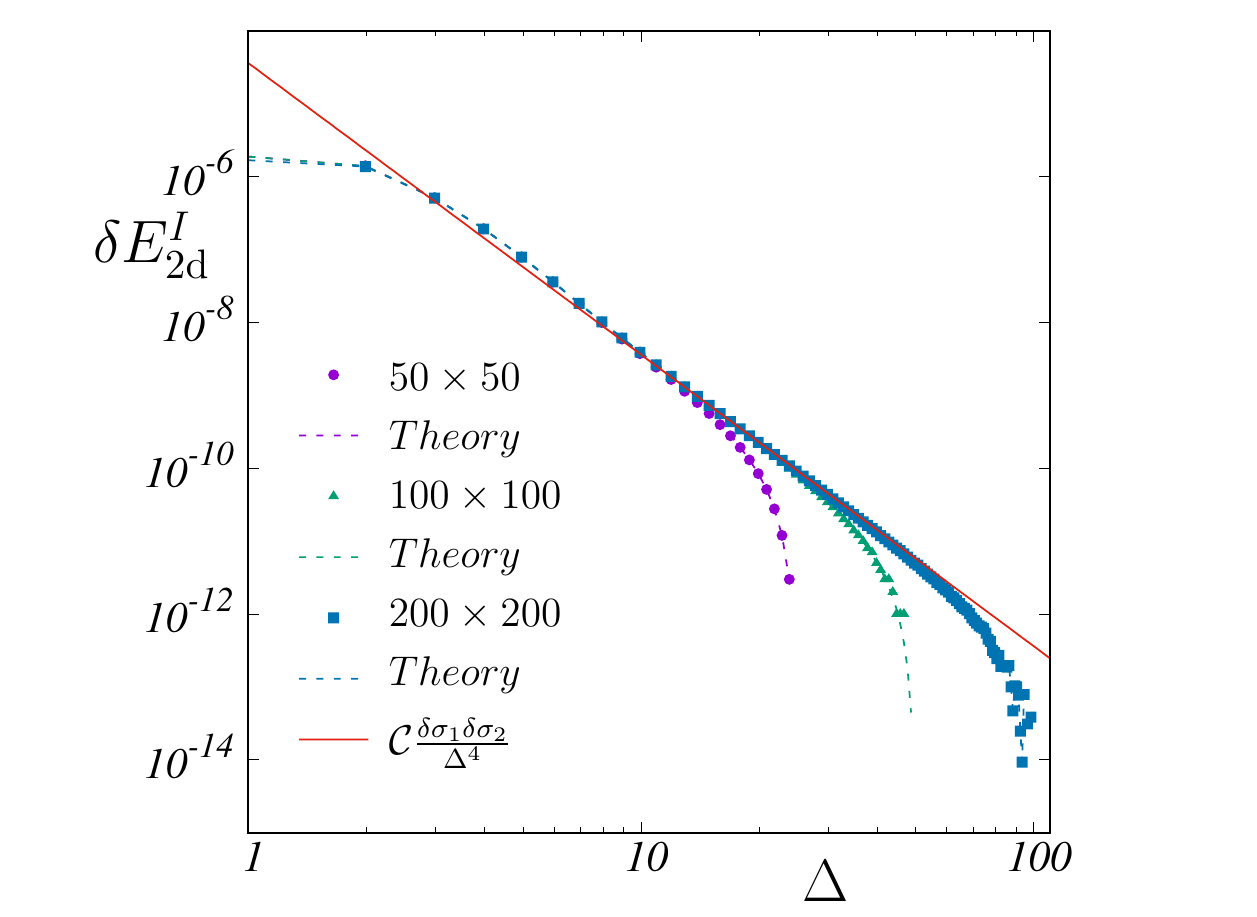}
\centering
\caption{Interaction energy between two defects placed at a distance $\Delta$ along the $x$-direction, along with the theoretical prediction in Eq.~(\ref{finite_ss_en_supp}) for three different system sizes. Here points represent the data from simulations while dashed lines represent the theoretical predictions. The solid red lines is the prediction for the interaction energy between two defects in the continuum limit, which displays a $\sim\Delta^{-4}$ behaviour.}
\label{finite_size_inten}
\end{figure}
\subsection{Interaction Energy between Defects}

In this Section, we examine the {\it interaction} energy between two defects $\delta E^I_{\text{2d}}$.
When two defects are present, the energy of the system can be expressed as a combination of several terms as
\begin{eqnarray}
E = E^{(0)} + \delta E_{\text{1d}}(\vec{r}_1) + \delta E_{\text{1d}}(\vec{r}_2) + \delta E^I_{\text{2d}}(\vec{r}_1,\vec{r}_2),
\end{eqnarray}
where $\delta E_{\text{1d}}(\vec{r}_1)$ and $\delta E_{\text{1d}}(\vec{r}_2)$ represent the excess energy associated with defects $1$ and $2$ placed at $\vec{r}_1$ and $\vec{r}_2$ respectively. $\delta E^I_{\text{2d}}(\vec{r}_1,\vec{r}_2)$ represents the energy of interaction between the defects. 
When the defects are separated by a large distance, the interaction energy tends to zero, and we have
\begin{eqnarray}
E_{\infty} = E^{(0)} + \delta E_{\text{1d}}(\vec{r}_1) + \delta E_{\text{1d}}(\vec{r}_2).
\end{eqnarray}   
Here $E_\infty$ is the energy of the system when the defects are located an infinite distance apart. From simulations of finite systems it is not possible to measure $E_\infty$, thus we choose $E_\infty=E_{\vec{L}_{\text{max}}}$ for an $L\times L$ system, where $\vec{L}_{\text{max}}$ represents the largest possible separation between defects along the chosen direction. Therefore in order to compute the energy of interaction from simulations, we calculate 
\begin{eqnarray}
\delta E^I_{\text{2d}}(\vec{r}_1,\vec{r}_2) &=& E - E_{\vec{L}_{\text{max}}}.
\label{eq_supp_finite_system_energy}
\end{eqnarray}

Without loss of generality, we place the first defect with incremental radius $\delta\sigma_1$ at the origin $\vec{0} = (0,0)$ and the second defect with $\delta\sigma_2$ at position $\vec{\Delta} = (\Delta_x,\Delta_y)$. Therefore $\delta \tilde{\sigma} (\vec{k})$ is given by
$\delta \tilde{\sigma} (\vec{k})= \sum_{\vec{r}}\delta \sigma \exp{(\mathrm{i}\vec{k}.\vec{r})}=
\delta\sigma_{1}+ \delta\sigma_{2}\exp{(\mathrm{i}\vec{k}.\vec{\Delta})}$.
The energy of interaction can then be expressed as
\begin{eqnarray}
\hspace{-0.4cm}
\delta E^I_{\text{2d}}(\vec{0},\vec{\Delta}) &=& \delta E_{\text{2d}}(\vec{0},\vec{\Delta}) - \delta E_{\text{1d}}(\vec{0}) - \delta E_{\text{1d}}(\vec{\Delta}),
\label{defdiff}
\end{eqnarray}
where $\delta E_{\text{2d}}(\vec{0},\vec{\Delta})$ is the excess energy of the system with defects placed at $(0,0)$ and $(\Delta_x,\Delta_y)$ together.

For ease of computation, we choose the two defects to be placed at a distance $\Delta$ apart along the $x$-axis. Here $\delta E_{\text{1d}}(\vec{0})$ and $\delta E_{\text{1d}}(\Delta)$ are the excess energies of the configuration with defects at $(0,0)$ and $(\Delta_x,\Delta_y)$ individually.
We therefore have 
\begin{eqnarray}
\nonumber
\delta E_{\text{1d}}(\vec{0}) &=& \frac{1}{2}\big(\gamma_1 \delta \tilde{\sigma}(0) + \delta\sigma_1^2\sum_{\vec{k}}\gamma_2(\vec{k})\big),\\
\nonumber
\delta E_{\text{1d}}(\vec{\Delta}) &=& \frac{1}{2}\big(\gamma_1 \delta \tilde{\sigma}(0) + \delta\sigma_2^2\sum_{\vec{k}}\gamma_2(\vec{k})\big),\\
\delta E_{\text{2d}}(\vec{0},\vec{\Delta})&=& \frac{1}{2}\Big(2\gamma_1 \delta \tilde{\sigma}(0)+ \gamma_2(\vec{k})\big(\delta\sigma_{1}+ \delta\sigma_{2}\exp{(\mathrm{i}\vec{k}.\vec{\Delta})}\big)\big(\delta\sigma_{1}+ \delta\sigma_{2}\exp{(-\mathrm{i}\vec{k}.\vec{\Delta})}\big)\Big).
\label{1def2def}
\end{eqnarray}
Using Eqs.~(\ref{defdiff}) and (\ref{1def2def}), the energy of interaction takes the form
\begin{eqnarray}
\langle \delta E^I_{\text{2d}}(\vec{0},\vec{\Delta}) \rangle &=&  \delta\sigma_1 \delta\sigma_2 \sum_{\vec{k}}\gamma_2(\vec{k})\cos(\vec{k}.\vec{\Delta})
\end{eqnarray}
For a finite system size using Eq.~(\ref{eq_supp_finite_system_energy}), the energy of interaction takes the following form
\begin{eqnarray}
\hspace{-0.0cm}
\nonumber
\langle \delta E^I_{\text{2d}}(\vec{0},\vec{\Delta}) \rangle  &=&  \delta\sigma_1 \delta\sigma_2 \sum_{\vec{k}}\gamma_2(\vec{k})\Big(\cos(\vec{k}.\vec{\Delta})-\cos \left(\vec{k}. \vec{L}_{\text{max}}\right)\Big),\\
\label{finite_ss_en_supp}
\end{eqnarray}
which is Eq.~(\ref{finite_ss_en}) in the main text. We plot the above expression compared with numerical simulations for three finite system sizes in Fig.~\ref{finite_size_inten}, showing an exact match between the two. In the continuum limit, the energy of interaction between two defects placed at a distance $\vec{\Delta}$ apart can be expressed as
\begin{eqnarray}
\langle \delta E^I_{\text{2d}}(\vec{0},\vec{\Delta}) \rangle &=& \frac{\delta\sigma_1 \delta\sigma_2}{4\pi^2}\int_{-\pi}^\pi\int_{-\pi}^\pi\gamma_2(\vec{k})\cos(\vec{k}.\vec{\Delta})dk_x dk_y.
\label{eq_inten}
\end{eqnarray}
In order to evaluate the above integration for two defects placed along the $x$-direction, we simplify the interaction kernel $\gamma_2(\vec{k})$. The exact form of the simplified $\gamma_2(\vec{k})$ provided in Eq.~(\ref{gamma2_exact}) can be expressed as
\begin{eqnarray}
\nonumber
\gamma_2(\vec{k})&=&\big.(\epsilon -1) \big(-864 \epsilon ^3 \cos ({k_x}) \cos ({k_y})+16 \epsilon ^3 \cos ({k_x}) \cos (3 {k_y})+1128 \epsilon ^2 \cos ({k_x}) \cos ({k_y})\\
\nonumber
&&-8 \epsilon ^2 \cos ({k_x}) \cos (3 {k_y})+\cos (4 {k_x}) \big(4 (1-2 \epsilon )^2 \epsilon  \cos ({k_x}) \cos ({k_y})
 +2 (1-2 \epsilon )^2 \epsilon  \cos (2 {k_y})\\
\nonumber
&&+80 \epsilon ^3-104 \epsilon ^2+35 \epsilon -3\big)+2 \cos (2 {k_x}) \big(-2 \cos ({k_x}) \big(4 \epsilon ^3 \cos (3 {k_y})\\
\nonumber
&&+\big(-20 \epsilon ^3+20 \epsilon ^2-23 \epsilon +6\big) \cos ({k_y})\big)+\big(80 \epsilon ^3-104 \epsilon ^2+35 \epsilon -3\big) \cos (2 {k_y})\\
\nonumber
&&-206 \epsilon ^3+272 \epsilon ^2-113 \epsilon +15\big)-498 \epsilon  \cos ({k_x}) \cos ({k_y})+2 \epsilon  \cos ({k_x}) \cos (3 {k_y})\\
\nonumber
&&+72 \cos ({k_x}) \cos ({k_y})-4 \epsilon ^3 \cos (6 {k_x})+24 \epsilon ^3 \cos (2 {k_y})-24 \epsilon ^2 \cos (2 {k_y})+24 \epsilon  \cos (2 {k_y})\\
\nonumber
&&-6 \cos (2 {k_y})+912 \epsilon ^3-1224 \epsilon ^2+495 \epsilon -63\big)\big)/16 \epsilon ^2 \cos ^2({k_x}) \cos (2 {k_y})\\
\nonumber
&&-96 \epsilon ^2 \cos ({k_x}) \cos ({k_y})+\cos (2 {k_x}) \big(4 \big(8 \epsilon ^2-8 \epsilon +3\big) \cos ({k_x}) \cos ({k_y})\\
\nonumber
&&-40 \epsilon ^2+40 \epsilon -9\big)-16 \epsilon  \cos ^2({k_x}) \cos (2 {k_y})+96 \epsilon  \cos ({k_x}) \cos ({k_y})-24 \cos ({k_x}) \cos ({k_y})\\
&&+4 (\epsilon -1) \epsilon  \cos (4 {k_x})+3 \cos (2 {k_y})+84 \epsilon ^2-84 \epsilon +18.
\label{gamma2}
\end{eqnarray}
Next, using the notation $\cos(k_x)\to X$ and $\cos(k_y)\to Y$, in Eq.~\eqref{eq_inten} and using De Moivre's formula further leads to
\begin{eqnarray}
\langle \delta E^I_{\text{2d}}(\vec{0},\vec{\Delta}) \rangle &=& \frac{4\delta\sigma_1 \delta\sigma_2}{4\pi^2}\int_{-1}^1\int_{-1}^1 \frac{f(X,Y,\epsilon)}{\sqrt{1-X^2}\sqrt{1-Y^2}}\big(X+\mathrm{i}\sqrt{1-X^2}\big)^\Delta dX dY.
\end{eqnarray}  
Here $f(X,Y,\epsilon) = \frac{f^n(X,Y,\epsilon)}{f^d(X,Y,\epsilon)}$ with $f^n(X,Y,\epsilon)$ and $f^d(X,Y,\epsilon)$ are the numerator and the denominator of the function $f(X,Y,\epsilon)$ respectively.  
 Next, factorizing the denominator of $f(X,Y,\epsilon)$ i.e. $f^d(X,Y,\epsilon)$, we can write
\begin{eqnarray}
\langle \delta E^I_{\text{2d}}(\vec{0},\vec{\Delta}) \rangle &=& \frac{\delta\sigma_1 \delta\sigma_2}{\pi^2}\int_{-1}^1\int_{-1}^1 \frac{f^n(X,Y,\epsilon)\left(X+\mathrm{i}\sqrt{1-X^2}\right)^\Delta dX dY}{C(X,\epsilon)\big(Y-Y_1(X,\epsilon)\big)\big(Y-Y_2(x,\epsilon)\big)\sqrt{1-X^2}\sqrt{1-Y^2}},
\end{eqnarray}
Using the method of partial fractions, the energy of interaction can next be expressed as
  \begin{eqnarray}
  \nonumber
  \langle \delta E^I_{\text{2d}}(\vec{0},\vec{\Delta}) \rangle &=& \frac{\delta\sigma_1 \delta\sigma_2}{\pi^2}\int_{-1}^1\int_{-1}^1 \frac{f^n(X,Y,\epsilon)\big(X+\mathrm{i}\sqrt{1-X^2}\big)^\Delta\big)dX dY}{\sqrt{1-X^2}\sqrt{1-Y^2}(4\sqrt{3}\sqrt{(-1+X^2)^3(-3+4\epsilon)(-1+4\epsilon)})}\\ 
&&  \Bigg(\frac{1}{\big(Y-Y_1(X,\epsilon)\big)}-\frac{1}{\big(Y-Y_2(X,\epsilon)\big)}\Bigg).
\end{eqnarray}   
   Evaluating the $Y$-integral yields
 \begin{eqnarray}
 \langle \delta E^I_{\text{2d}}(\vec{0},\vec{\Delta}) \rangle &=& \frac{\delta\sigma_1 \delta\sigma_2}{\pi^2}\int_{-1}^1 F(X,\epsilon)\big(X+\mathrm{i}\sqrt{1-X^2}\big)^\Delta dx
\end{eqnarray} 
The function $F(X,\epsilon)$ is symmetric under the transformation $X\to-X$, thus it is convenient to expand this function in terms of $X^2$. Now as the major contribution of the integration comes from $X^2\to1$, we can expand the function around $1$. The function $F(X,\epsilon)$ takes the form $F(u,\epsilon)$ with the transformation $X^2-1\to u$. Thus, the above described expansion is equivalent to expanding the function $F(u,\epsilon)$ around $u=0$. Expanding the function we get

\begin{eqnarray}
F(u,\epsilon)&=& F_{\left(-\frac{1}{2}\right)}(\epsilon) u^{-1/2} + F_{\left(\frac{1}{2}\right)}(\epsilon) u^{1/2} + F_{\left(1 \right)}(\epsilon) u^{1} + F_{\left(\frac{3}{2}\right)}(\epsilon) u^{3/2} + F_{\left(2\right)}(\epsilon) u^{2} + \mathcal{O}(u^\frac{5}{2}).
\end{eqnarray}

The above coefficients are functions of $\epsilon$ and have the following form,
\begin{eqnarray}
\nonumber
F_{\left(-\frac{1}{2}\right)}(\epsilon) &=& -\frac{16 \mathrm{i} \pi  \left(\epsilon -1 \right) \left(2 \epsilon  \left (6 \epsilon -7 \right )+3 \right)}{\left(4 \epsilon -3\right)}, \\
\nonumber
F_{\left(\frac{1}{2}\right)}(\epsilon) &=& -\frac{16 \mathrm{i} \pi  \left(\epsilon -1\right) \left( \epsilon  \left (4 \epsilon \left (8 \epsilon  \left (2 \epsilon -7 \right)+53 \right )-75 \right )+9 \right )}{\left (3-4 \epsilon \right)^2 \left(4 \epsilon -1 \right)}, \\
\nonumber
F_{\left(1 \right)}(\epsilon) &=& \frac{128 \sqrt{3} \pi \left (1-2 \epsilon \right )^2 \left (\epsilon -1 \right )}{\left (3-4 \epsilon \right)^2}, \\
\nonumber
F_{\left(\frac{3}{2} \right)}(\epsilon) &=& -\frac{256  \mathrm{i} \pi  (1-2 \epsilon )^2 (\epsilon -1) \epsilon  \left(4 \epsilon  \left(8 \epsilon ^2+\epsilon -9\right)+9\right)}{(1-4 \epsilon )^2 (4 \epsilon -3)^3}, \\
F_{\left(2 \right)}(\epsilon) &=& -\frac{64 \left(\pi  (1-2 \epsilon )^2 (\epsilon -1) \left(64 \epsilon ^3-48 \epsilon -17\right)\right)}{\sqrt{3} (4 \epsilon -3)^3 (4 \epsilon -1)}.
\end{eqnarray}
Putting back the value of $u=X^2-1$, and using the substitution $X = \text{cos}(k_x)$, the energy of interaction takes the form

\begin{eqnarray}
\langle \delta E^I_{\text{2d}}(\vec{0},\vec{\Delta}) \rangle &=& \frac{\delta\sigma_1 \delta\sigma_2}{\pi^2} \Bigg[ \int_{0}^\pi F_{(-\frac{1}{2})}(\epsilon)\text{exp}(\mathrm{i} k_x \Delta) d k_x +
\int_{0}^\pi F_{(\frac{1}{2})}(\epsilon) \text{sin}^2 (k_x) \text{exp}(\mathrm{i} k_x \Delta) d k_x \\
 \nonumber
&& + \int_{0}^\pi F_{({1})}(\epsilon)\text{sin}^3 (k_x) \text{exp}(\mathrm{i} k_x \Delta) d k_x + \int_{0}^\pi F_{(\frac{3}{2})}(\epsilon)\text{sin}^4 (k_x) \text{exp}(\mathrm{i} k_x \Delta) d k_x \\
\nonumber
&& + \int_{0}^\pi F_{({2})} (\epsilon)\text{sin}^5 (k_x) \text{exp}(\mathrm{i} k_x \Delta) d k_x + \mathcal{O}\big(\text{sin}^6 (k_x) \big) \Bigg] 
\end{eqnarray}
Due to the lattice notation we choose, $\Delta$ is always even (see Refs.~\cite{acharya2020athermal,das2021long,horiguchi1972lattice}). This makes the first two terms in the expression exactly zero. To extract the dependence on $\Delta$ in the limit $\Delta \to \infty$, we use the following identity:
\begin{eqnarray}
\nonumber
\int^{\pi}_{0} \text{sin}^{2m+1} (\xi) \text{exp} ({\mathrm{i}} \Delta \xi) d \xi &=& 
\frac{1}{2^{2m -1 }}\sum_{k =0}^{m} \frac{(-1)^k(2k + 1)}{(2m +1)^2 - \Delta^2}\binom{2m+1}{m-k} \\
&\sim & \frac{(2m+1)!}{2 \Delta^{2m+2}} \hspace{1.1 in} (\Delta \to \infty).
\end{eqnarray}
Thus, the interaction energy in the limit $\Delta \to \infty$ is
\begin{equation}
\langle \delta E^I_{\text{2d}}(\vec{0},\vec{\Delta}) \rangle \simeq \frac{16 \delta \sigma^2}{\pi} \frac{24 \sqrt{3} (1-2 \epsilon )^2 (\epsilon -1)}{(3-4 \epsilon )^2 \Delta^4}.
\label{en_form}
\end{equation}
The above equation contains $\Delta$ in index space as described in Ref.~\cite{horiguchi1972lattice}. 
Finally, the energy of interaction in Eq.~(\ref{en_form}) can then be written in real space as
\begin{equation}
\langle \delta E^I_{\text{2d}}(\vec{0},\vec{\Delta}) \rangle \simeq  \frac{ \delta\sigma_1 \delta\sigma_2}{\pi} \frac{24 \sqrt{3} (1-2 \epsilon )^2 (\epsilon -1)}{(3-4 \epsilon )^2 \Delta ^4 },
\end{equation}
which is Eq.~(\ref{powerlaw}) in the main text. We plot this expression along with the interaction energy obtained from numerical simulations when two defects are placed along $x$-direction in Fig.~\ref{finite_size_inten}. The simulation and theory match perfectly and yield a $\Delta^{-4}$ behaviour at large separations.

\subsection{Fluctuations in Interaction Energy}
In order to calculate the fluctuations in the interaction energy between defects due to the underlying disorder, we separate the contributions of the defects, the disorder and the crystalline background. The total energy of a disordered system with two defects separated by a distance $\vec{\Delta}$ can be expressed as
\begin{eqnarray}
E(\Delta) &=& E^{(0)} + \frac{1}{2}\gamma_1(\vec{0})\Big(\delta\tilde{\sigma}_{\text{dis}}(\vec{0})+\delta\tilde{\sigma}_{\text{def}}(\vec{0})\Big) + \frac{1}{2}\sum_{\vec{k}}\gamma_2(\vec{k})\Big(\delta\tilde{\sigma}_{\text{dis}}(\vec{k})+\delta\tilde{\sigma}_{\text{def}}(\vec{k})\Big)\Big(\delta\tilde{\sigma}_{\text{dis}}(-\vec{k})+\delta\tilde{\sigma}_{\text{def}}(-\vec{k})\Big).
\end{eqnarray}
Omitting the contributions arising from disorder and the individual defect alone, the interaction energy between two defects in a disordered background for an infinite system size can be expressed as
\begin{eqnarray}
\delta E_{\text{2d}}^I &=& \frac{1}{2}\sum_{\vec{k}}\gamma_2(\vec{k})(\delta\tilde{\sigma}_{\text{def}}(-\vec{k})\delta\tilde{\sigma}_{\text{dis}}(\vec{k})+\delta\tilde{\sigma}_{\text{def}}(\vec{k})\delta\tilde{\sigma}_{\text{dis}}(-\vec{k})+\delta\tilde{\sigma}_{\text{def}}(-\vec{k})\delta\tilde{\sigma}_{\text{def}}(\vec{k})).
\end{eqnarray}
We note that the third term on the right hand side of the above equation is the contribution arising from the interaction between defects on a crystalline background.  Taking the third term to the left hand side, the equation can be re-expressed as
\begin{eqnarray}
\delta E_{\text{2d}}^I-\langle\delta E_{\text{2d}}^I\rangle &=& \frac{1}{2}\sum_{\vec{k}}\gamma_2(\vec{k})(\delta\tilde{\sigma}_{\text{def}}(-\vec{k})\delta\tilde{\sigma}_{\text{dis}}(\vec{k})+\delta\tilde{\sigma}_{\text{def}}(\vec{k})\delta \tilde{\sigma}_{\text{dis}}(-\vec{k})).
\end{eqnarray}
Therefore, the variance of the interaction energy has the following form 
\begin{eqnarray}
\langle(\delta E_{\text{2d}}^I-\langle\delta E_{\text{2d}}^I\rangle)^2\rangle &=&\frac{V\eta^2}{48}\sum_{\vec{k}}
 \gamma_2(\vec{k})\gamma_2(-\vec{k})\left(\delta\sigma_1^2+\delta\sigma_2^2+\delta\sigma_1\delta\sigma_2 \cos(\vec{k}.\vec{\Delta})\right),
\end{eqnarray}
which is Eq.~(\ref{eq_interaction_energy_fluctuation_main}) in the main text.

\clearpage
\end{widetext}
\end{document}